\documentclass[reprint,prb,aps,amsfonts,amssymb,amsmath,floatfix,showpacs]{revtex4-1}
\usepackage[]{graphicx}
\usepackage{epstopdf}
\usepackage{physics}
\usepackage{array}
\usepackage{booktabs}
\usepackage{multirow}
\usepackage{gensymb}
\usepackage{dcolumn}

\newcommand{\etal}{\textit{et al.}}

\begin{document}
\title{Nonlinear optical responses of all-inorganic lead halide perovskite nanostructures by time-resolved beam-deflection technique}
\author{Samrat Roy}
\affiliation{Department of Physical Sciences, Indian Institute of Science Education and Research Kolkata, Mohanpur, Nadia 741246, West Bengal, India}
\author{Arnab Mandal}
\affiliation{Department of Chemical Sciences and Centre for Advanced Functional Materials, Indian Institute of Science Education and Research Kolkata, Mohanpur, Nadia 741246, West Bengal, India}
\author{Akshay Raj R.}
\affiliation{Department of Physical Sciences, Indian Institute of Science Education and Research Kolkata, Mohanpur, Nadia 741246, West Bengal, India}
\author{Sayan Bhattacharyya}
\affiliation{Department of Chemical Sciences and Centre for Advanced Functional Materials, Indian Institute of Science Education and Research Kolkata, Mohanpur, Nadia 741246, West Bengal, India}
\author{Bipul Pal}
\email{bipul@iiserkol.ac.in}
\affiliation{Department of Physical Sciences, Indian Institute of Science Education and Research Kolkata, Mohanpur, Nadia 741246, West Bengal, India}
\date{\today}
\preprint{To be submitted to .......}

\begin{abstract}
We have investigated nonlinear refraction in all-inorganic halide perovskites, the CsPbBr$_3$ and CsPbBr$_{1.5}$I$_{1.5}$ nanosheet and quantum dot colloids in toluene, by a novel beam deflection technique using near-resonant continuous wave lasers. The nonlinear refraction and its time-evolution measured here have originated from thermal lensing effect. Nonlinear behaviour of heat transport in terms of intensity dependent thermal diffusion rate has been observed. Effects of convective heat flow have been measured at high intensities. Quantum dots have higher nonlinear refraction as compared to the corresponding nanosheet samples, presumably due to reduced dimensionality. The effective values of nonlinear refractive index, estimated here for near-resonant excitations, exceed those reported in the literature for organic-inorganic hybrid perovskites in the nonresonant excitation regime, by several orders of magnitude. 
\end{abstract}

\maketitle

\section{Introduction}
Materials showing interesting optical responses have always attracted the attention of mankind in general and researchers in particular.~\cite{Weber2018, Nikogosyan1997} A variety of new direct band semiconductors have populated the plethora of optoelectronic materials in recent times.~\cite{Mak2016, Bhimanapati2015, Lim2011} Notable among them are the organic-inorganic hybrid metal-halide perovskites with state-of-the-art light-matter interactions.~\cite{Sutherland2016, Song2015} General formula for a class of highly luminescent perovskite semiconductors is given by ABX$_3$ where A is usually a monovalent organic cation, B is a divalent metal ion, and X is a halide anion.~\cite{Saparov2016, You2015} Among the most well-known optically active perovskites are those where A is either caesium, Cs$^+$, methylammonium, CH$_3$NH${_3}{^+}$ (MA$^+$), or formamidinium, HC(NH$_2$)$_2^+$ (FA$^+$); B is Pb$^{2+}$; and X is I$^-$, Br$^-$, or Cl$^-$. They have emerged as promising semiconductors due to their unique properties, such as broad chemical tunability,~\cite{Meloni2016, You2014, Habisreutinger2016} excellent charge transport properties, limited charge recombination, higher diffusion length, high defect tolerance and so on.~\cite{Xing2013, Wehrenfennig2013, Ponseca2014, Dong2015, Gao2014, Tan2014} Due to the stability issues of organic-inorganic hybrid perovskites, all inorganic perovskites,~\cite{Shirayama2016, Swarnkar2016} such as  CsPbX$_3$ (X~$=$~Cl, Br, I, or mixture of these) emerge as suitable alternatives, owing to their better stability and excellent optoelectronic properties.~\cite{Yakunin2015, Protesescu2015, Akkerman2015} Tunable band gap of CsPbX$_3$ for different halide compositions essentially indicates that there is a variation in light-matter interaction.~\cite{Filip2014, Lee2017} These interactions may also be controlled by tuning the dimensionality, from bulk (3D) to sheet (2D) to rod (1D) to dot (0D), through proper choice of reaction conditions.~\cite{Zhang2015, Sun2016, Liang2016, Shamsi2016, Zhang2016a} Colloidal quantum dots (QDs) of all inorganic lead halide perovskites represent the latest entries with solution processability, showing high quantum yield, narrow emission linewidth, and tunable absorption and photoluminescence (PL) spectra.\cite{Ghosh2018, Swarnkar2016, Yakunin2015, Nedelcu2015} Two-dimensional nanosheets (NSs) of these materials offer unique advantages over their 3D counterparts in terms of increased specific surface area, narrow full-width at half-maximum (FWHM) of PL spectra, and shorter exciton lifetime, which make them a good candidate for light emitting applications.~\cite{Li2017, Wang2017, Song2015, Yang2018, Song2017} Along with high power conversion efficiency, high PL quantum yield in the visible range, long carrier diffusion length, and small exciton binding energy makes these materials important for solar cell applications.~\cite{Meloni2016, Chen2015, You2015, Eperon2015, Ghosh2019, Bera2019, Tai2019, Wang2019, Zhang2018} 

On the other hand, for the development of high speed optical switching, lasers, and photonic devices, understanding of nonlinear refraction and nonlinear absorption under intense laser illumination is essential for any photo sensitive material.~\cite{2017} Although there are several reports on nonlinear optical effects in organic-inorganic mixed halide perovskites, for CsPbX$_3$, much attention has been focused on  their synthesis and photovoltaic applications.~\cite{Kalanoor2016, Lu2016, Zhang2016, Yi2017, Zheng2019, Duan2019, Kostopoulou2018, Dong2017} There are very few reports on their nonlinear optical behaviour and applications, except a few based on the thinfilm form of these materials.~\cite{Wang2015, Wu2018, Krishnakanth2018} Although, near-resonant excitation may enhance nonlinear optical responses, available reports of nonlinear optical measurements on perovskite nanostructures have mainly used off-resonant excitation with very high power (amplified) femtosecond pulsed lasers, which makes it difficult for mass-scale applications outside laboratories. 

Here, we study the nonlinear optical properties of CsPbX$_3$ (X~$=$~Br and Br:I in 50:50 stoichiometry) QD and NS colloids in toluene by time-resolved beam-deflection technique using continuous wave (CW) lasers. In our two-colour pump-probe experimental scheme, the pump beam energy was set above the absorption edges of our samples, to exploit the possible resonant enhancement of optical nonlinearities. Nonlinear refraction may have different origins. One of them is due to the instantaneous third order nonlinear polarization arising under intense laser excitation, called third order $\chi^{(3)}$ processes.~\cite{Sheik-Bahae1990} This arises when strong electric field of the incident intense laser light significantly modifies the Coulomb potential between the atomic electrons and the nucleus.~\cite{Boyd2008, Leuthold2010, Sutherland2016} Another origin of nonlinear refraction is due to the thermally stimulated density variation which results from laser heating and thermal expansion of the sample. The latter one is the dominant process for CW laser excitation, especially in liquid samples or samples dispersed in liquid medium. This is popularly known as the thermal lensing effect.~\cite{Dabby1968, Leite1967, Maity2012} The radially varying intensity profile of a Gaussian laser beam sets up a radially varying temperature profile in the sample. This results in a radial density gradient leading to a refractive index gradient over the beam cross-section. As a result, the sample acts like a lens. The formation dynamics of thermal lens depends on several physical properties, such as the thermal conductivity, specific heat, density and its temperature coefficient, temperature coefficient of refractive index, bulk modulus, viscosity, etc.~\cite{Sheldon1982} Many of these properties can be investigated, in a contact-less optical method by time-resolved study of the thermal lensing effect. Thermal lensing effect may be effectively described by an intensity-dependent refractive index $n = n_0+n_2I(r)$,~\cite{Boyd2008, Sutherland2016} where $n_0$ is the linear refractive index, $n_2$ is the so-called  nonlinear refractive index, and $I(r)$ is the laser intensity profile. Here we have assumed thin sample approximation where variation of $I(r)$ within the sample thickness along the beam propagation direction is neglected. Our experiments yield good estimates of the effective $n_2$ values in CsPbX$_3$ (X~$=$~Br and 50:50 stoichiometry of Br:I) colloidal QDs and NSs in toluene (1~mg/ml).
 
\section{Samples, synthesis and characterization}
Our study is based on the following samples: (a) CsPbBr$_3$ QDs, designated as S1, (b) CsPbBr$_3$ NSs, designated as S2, (c) CsPbBr$_{1.5}$I$_{1.5}$ QDs, designated as S3, and (d) CsPbBr$_{1.5}$I$_{1.5}$ NSs, designated as S4. Due to the large bandgap (lying in the deep UV region) of CsPbCl$_3$ and  stability issues of CsPbI$_3$ in ambient air, those samples were not studied here.

The QDs were synthesized by following the method described by Protesescu {\etal}~\cite{Protesescu2015} and Ghosh {\etal}~\cite{Ghosh2018} For the synthesis of colloidal QDs, oleic acid and oleyl amine were used as surfactants. The NSs were  prepared by following the work of Shamsi {\etal}~\cite{Shamsi2016} with slight modifications. The samples were characterized by transmission electron microscopy (TEM; JEOL, JEM-2100F micro-scope using 200kV electron source at the DST-FIST facility, IISER Kolkata), atomic force microscopy (AFM; NT-MDT NTEGRA instrument from NT-MDT, Santa clara, USA), X-ray diffraction (XRD; Rigaku mini flex II, Japan with Cu K$\alpha$ radiation of 1.54054~{\AA}  wavelength), UV-visible absorption (JASCO V-670 spectrophotometer) and PL spectroscopy (Acton SP2500 spectrograph fitted with a Princeton PIXIS-100 CCD).

\section{Beam-deflection technique}
\begin{figure}[htb]
    \includegraphics[clip,width=0.8\linewidth]{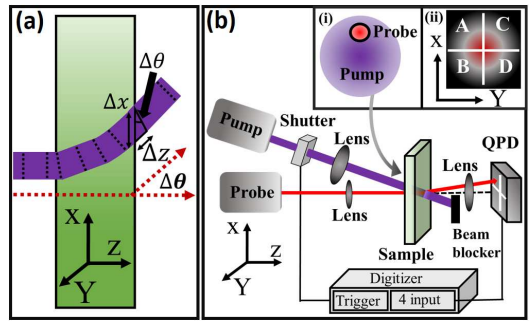}
    \caption{Schematics of (a) beam bending in a medium having a transverse, 1D refractive index gradient, (b) the beam-deflection setup, (i) blow-up of the pump-probe overlap region, and (ii) front-view of the quadrant photodiode (QPD).} \label{setup}
\end{figure}

The nonlinear optical properties of the QDs and NSs were studied by irradiating with an intense Gaussian laser beam whose energy was chosen above the absorption edges of the samples, to enable strong absorption of light, giving rise to a radially varying refractive index profile through thermal lensing effect. We employ the idea that the path of a laser beam propagating through a medium, having a refractive index gradient in the transverse direction, will bend because different parts of the wavefront will travel with different speeds, as they experience different values of refractive index. This is schematically shown in Fig.~\ref{setup}(a). A calculation of the beam bending angle per unit length of propagation, say in $z$-direction, within the sample having a one-dimensional refractive index gradient, say in $x$-direction, under thin medium approximation and considering absence of any intensity dependent nonlinear absorption gives~\cite{Casperson1973}
\begin{equation}
\dv{\theta}{z} = - \frac{1}{n_0} \dv{n}{x}, \label{angular bending}
\end{equation}
where $n_0$ is the average value and $\dv*{n}{x}$ is the gradient of the refractive index. 

Based on this idea we developed a pump-probe beam deflection experiment for time-resolved study of thermal lensing effect using CW lasers [Fig.~\ref{setup}(b)]. A near-resonant pump beam from a CW diode laser (wavelength $\lambda=403$~nm) starts to excite the samples (colloidal QDs and NSs of CsPbBr$_3$ and CsPbBr$_{1.5}$I$_{1.5}$ in toluene) taken in a 2~mm quartz cuvette, when the computer controlled shutter in the pump beam path is opened. As the pump laser beam heats up the sample, a thermal lens (radial gradient of refractive index) begins to develop. An off-resonant probe beam (energy lying below the absorption edges of the samples) from another CW diode laser ($\lambda=791$~nm) falling off-centred with respect to the pump beam spot [Fig.~\ref{setup}(b)(i)], gets deflected due to the refractive index gradient created by the pump beam. Although the pump and probe beams were co-focused on to the sample surface, the focusing lenses of the pump and probe beams were adjusted such that the pump beam was loosely focused (spot size $\approx 150$~$\mu$m) while the probe beam was tightly focused (spot size $\approx 30$~$\mu$m). The probe beam was set along the normal to the sample cuvette whereas the pump beam makes an angle $\lesssim 10\degree$. The probe beam position was varied within the pump beam spot to maximize the probe deflection. It was seen using a CCD camera that the probe deflection was maximum when it was placed near the half-width position of the pump beam spot, as expected from a simple theoretical calculation, discussed shortly. The deflection of the probe beam was measured by a quadrant photodiode (QPD) and recorded as a function of time, starting from opening the pump-beam shutter, by a 16-bit digitizer synchronized with the shutter controller. The digitizer is capable of recording data at 20 MHz which means an achievable time resolution better than $0.1$~$\mu$s. Actual time resolution of about 20~$\mu$s achieved in our experiments is limited by the shutter speed. The position coordinate $(x, y)$ of the laser beam spot on the QPD is given in an arbitrary unit~\cite{Li2018, Li2019} by $x = [(V_A+V_C)-(V_B+V_D)]/[V_A+V_B+V_C+V_D]$ and $y = [(V_A+V_B)-(V_C+V_D)]/[V_A+V_B+V_C+V_D]$, where $V_A, V_B, V_C$ and $V_D$ are the output voltages, respectively, of the A, B, C and D photodiodes of the QPD [Fig.~\ref{setup}(b)(ii)]. The position coordinates of the probe beam spot on the QPD, $(x_0, y_0)$ in absence of the pump beam, and $(x_1, y_1)$ in presence of the pump beam were measured and the deflection of the probe beam was calculated as $\Delta l = \sqrt{(x_1-x_0)^2+(y_1-y_0)^2} \times \gamma$~mm. The scaling factor $\gamma = 5.97$~mm was obtained through calibration with actual deflection of the beam in a controlled experiment. Finally, the angular deflection $\Delta \theta$ of the probe beam was calculated as $\Delta l/f$ where $f$ is the focal length of the lens used after the sample cuvette for collimating the probe beam.  

The transverse intensity profile of a Gaussian laser beam, $I(r) = I_0 \exp [-2(r/w_e)^2]$, $w_e$ is the beam waist ($1/e^2$-radius), results in radially varying refractive index through $n(r) = n_0+n_2 I(r)$. For simplicity, we consider only 1D refractive index variation, say, in $x$-direction. Total accumulated deflection for a probe beam, after propagating a distance $L$ through the pump beam generated thermal lens, can be calculated using Eq.~\eqref{angular bending} as~\cite{Ferdinandus2013}
\begin{equation}
\Delta \theta = \int_{z=0}^L \frac{1}{n_0} \dv{n}{x} \dd{z}. \label{total bending}
\end{equation}
Assuming negligible absorption, we take $\dv*{n}{x}$ to be independent of $z$ within the propagation length $L$ through the sample to get  $\Delta \theta=(n_2/n_0) L\dv*{I}{x}$. We can easily infer that the probe beam deflection will be maximum if it is placed at the point of inflection of the pump beam intensity profile. Setting $x=0.5 w_e$ and $y= 0$, we get the maximum deflection angle as 
\begin{equation}
\Delta \theta_m = \frac{4}{\pi\sqrt{e}} \frac{L}{w_e^3} n_2 P_\text{av}, \label{max deflection}
\end{equation}
where the average power of the pump beam, $P_\text{av}$, measured in the experiment, is related to the peak intensity $I_0$ and the waist size $w_e$ of the pump beam as $2 P_\text{av} = I_0 \pi w_e^2$.  The factor $n_0$ appearing in the denominator of Eq.~\eqref{total bending} is cancelled in Eq.~\eqref{max deflection} when refraction at the sample-air boundary in the exit plane is considered. In the input plane, the probe beam is normal to the sample surface. Equation~\eqref{max deflection} suggests that the plot of the maximum deflection angle as a function of the average power of the pump laser beam will be a straight line and the nonlinear refractive index $n_2$ can be calculated from its slope.

\section{Results and discussion}

\begin{figure}[htb]
    \includegraphics[clip,width=0.8\linewidth]{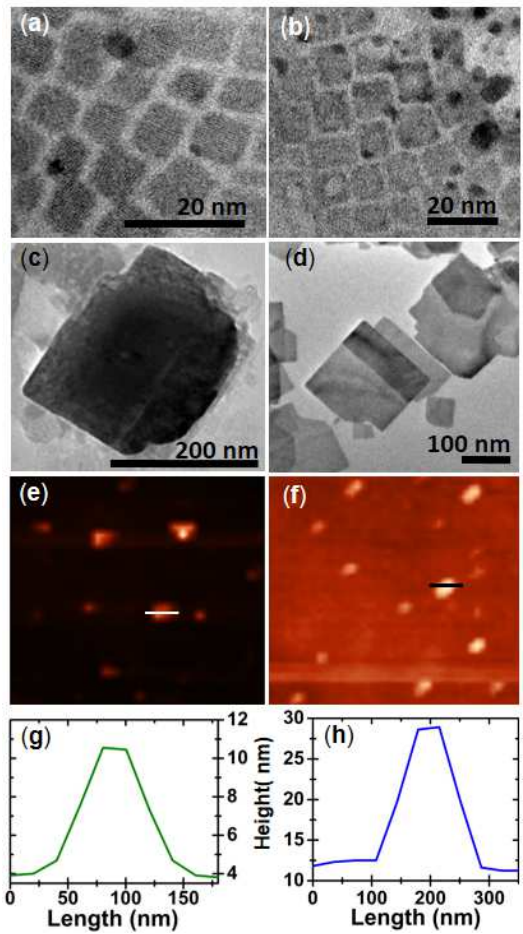}
    \caption{TEM images of (a) CsPbBr$_3$ QDs, (b) CsPbBr$_{1.5}$I$_{1.5}$ QDs, (c) CsPbBr$_3$ NSs, (d) CsPbBr$_{1.5}$I$_{1.5}$ NSs. The AFM images of (e) CsPbBr$_3$ NSs, and (f) CsPbBr$_{1.5}$I$_{1.5}$ NSs. The height profiles obtained from AFM scans for (g) CsPbBr$_3$ NSs, and (h) CsPbBr$_{1.5}$I$_{1.5}$ NSs.}\label{afm_tem}
\end{figure}

The TEM images in Figs.~\ref{afm_tem}(a,b) show the average edge-lengths of the CsPbBr$_3$ and CsPbBr$_{1.5}$I$_{1.5}$ QDs (nanocubes) to be $7.23 \pm 0.61$ and  $10.23 \pm 0.98$~nm, respectively. The lateral dimensions of the CsPbBr$_3$ and CsPbBr$_{1.5}$I$_{1.5}$ NSs are $153.82 \pm 6.92$ and $204.47 \pm 18.8$~nm, respectively [Figs.~\ref{afm_tem}(c,d)]. The height profiles obtained from the AFM scans [Figs.~\ref{afm_tem}(e,f)] provide the thickness of the CsPbBr$_3$ and CsPbBr$_{1.5}$I$_{1.5}$ NSs to be $7.04 \pm 0.65$ and $19.20 \pm 2.16$~nm, respectively [Figs.~\ref{afm_tem}(g,h)].

\begin{figure}[htb]
    \includegraphics[clip,width=0.8\linewidth]{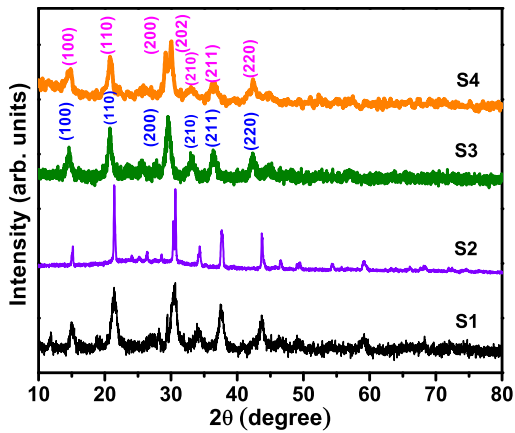}
    \caption{XRD patterns of S1 (CsPbBr$_3$ QDs), S2 (CsPbBr$_3$ NSs), S3 (CsPbBr$_{1.5}$I$_{1.5}$ QDs), and S4 (CsPbBr$_{1.5}$I$_{1.5}$ NSs).}\label{xrd}
\end{figure}

The XRD peaks of CsPbBr$_{1.5}$I$_{1.5}$ shift slightly to lower angle in comparison with CsPbBr$_3$  for both QDs and NSs [Fig.~\ref{xrd}]. This indicates the larger lattice spacing of CsPbBr$_{1.5}$I$_{1.5}$. The bifurcation into (200) and (202) reflections near $2\theta =30\degree$ for the NS samples, indicates their crystallization predominantly into the orthorhombic phase according to the space group \textit{Pbnm}.

\begin{figure}[htb]
\includegraphics[clip,width=0.8\linewidth]{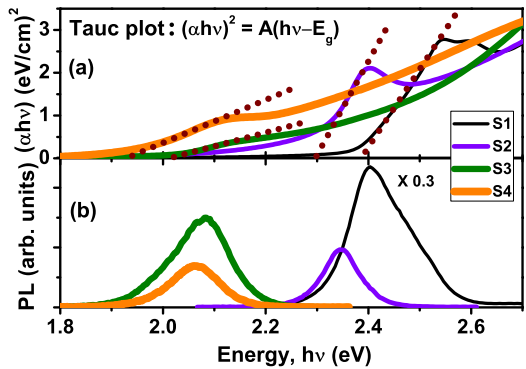}
\caption{(a) Tauc plots constructed from absorption spectra ($\alpha =$~absorption coefficient) and (b) PL spectra for S1 (CsPbBr$_3$ QDs), S2 (CsPbBr$_3$ NSs), S3 (CsPbBr$_{1.5}$I$_{1.5}$ QDs), and S4 (CsPbBr$_{1.5}$I$_{1.5}$ NSs). The linear fits (doted lines) in (a) are extrapolated to estimate the bandgap energies.}\label{tauc_pl}
\end{figure}

The Tauc plots in Fig.~\ref{tauc_pl}(a), constructed from the measured UV-visible absorption spectra, represent the relation $\alpha h \nu=A(h \nu - E_g)^n$, where $\alpha =$~absorption coefficient, $h\nu =$~photon energy, $E_g =$~optical bandgap of the material, $A$ is a constant, and the exponent $n$ depends on the nature of the optical transition. For direct allowed transitions, as in the case of CsPbX$_3$, $n=1/2$. Linear portions of each of the graphs are fitted with straight lines which are extrapolated to meet the abscissa ($\alpha h \nu=0$) at points giving $E_g=h\nu$. The values of $E_g$ thus obtained are given in Table~\ref{comparison}. 

\begin{table}[htb]
    \caption{Comparison of the bandgap, Stokes' shift, and PL FWHM for the QD and NS samples.}
    \label{comparison}
    \begin{tabular}{|c|c|c|c|c|}
        \hline
        \multicolumn{2}{|c|}{\multirow{2}{*}{Sample}} & Bandgap & Stokes' shift & PL FWHM \\ 
        \multicolumn{2}{|c|}{} & (eV) & (meV) & (meV) \\ \hline
        \multirow{2}{*}{QDs} & CsPbBr$_3$ & $2.40 \pm 0.05$ & $125 \pm 5$ & $14.0 \pm 0.2$ \\ \cline{2-5} 
        & CsPbBr$_{1.5}$I$_{1.5}$ & $2.02 \pm 0.02$ & $55 \pm 2$ & $13.4 \pm 0.2$ \\ \hline
        \multirow{2}{*}{NSs} & CsPbBr$_3$ & $2.29 \pm 0.01$ & $52 \pm 1$ & $8.7 \pm 0.1$ \\ \cline{2-5} 
        & CsPbBr$_{1.5}$I$_{1.5}$ & $1.92 \pm 0.02$ & $51 \pm 2$ & $10.9 \pm 0.1$ \\ \hline
    \end{tabular}
\end{table}

The PL spectra measured at room temperature under excitation at $\lambda = 403$~nm is presented in Fig.~\ref{tauc_pl}(b). The QDs are giving brighter PL (especially, CsPbBr$_3$ QDs are very bright) as compared to the corresponding NSs. This is expected because tighter quantum confinement leads to higher oscillator strength  in QDs. Stokes' shift between the absorption and PL peaks, and the PL linewidths, which are indicative of the optical quality of the samples, are listed in Table~\ref{comparison}. Somewhat wider PL spectrum of CsPbBr$_{1.5}$I$_{1.5}$ NSs as compared to that of the CsPbBr$_3$ NSs may arise from the mixed halide composition. Weaker PL intensity in CsPbBr$_{1.5}$I$_{1.5}$ as compared to CsPbBr$_3$ samples is indicative of reduced radiative efficiency which also may be caused by the Br-I composition modulations. Compared to the NSs, QDs have slightly larger PL linewidths, since similar fluctuations in size would lead to higher fluctuations in energy for the states having higher energy. 
             
\begin{figure}[htb]
	\includegraphics[clip,width=0.8\linewidth]{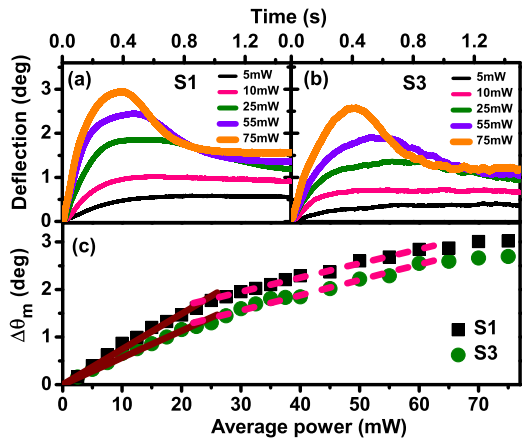}
	\caption{Time evolution of the probe beam deflection angle at a few selected pump powers for (a) CsPbBr$_3$ and (b) CsPbBr$_{1.5}$I$_{1.5}$ QDs. (c) Maximum deflection angle, $\Delta \theta_m$, as a function of the pump power. Data in the low and moderate power regimes are fitted with straight lines of slightly different slopes. Saturation is observed at high powers.}
	\label{qd_timeevolution}
\end{figure}

\begin{figure}[htb]
    \includegraphics[clip,width=0.8\linewidth]{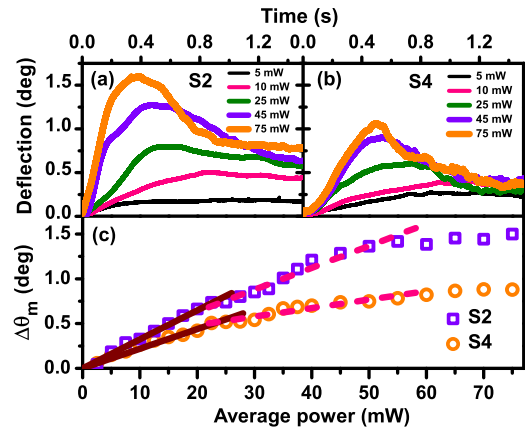}
    \caption{Time evolution of the probe beam deflection angle at a few selected pump powers for (a) CsPbBr$_3$ and (b) CsPbBr$_{1.5}$I$_{1.5}$ NSs (c) Maximum deflection angle, $\Delta \theta_m$, as a function of the pump power. Data in the low and moderate power regimes are fitted with straight lines of slightly different slopes. Saturation is observed at high powers.}
    \label{ns_timeevolution}
\end{figure}

Time-resolved pump-probe beam deflection measurements were performed under ambient temperature and pressure for different pump powers between 1--75~mW in  0--2~s time window. Time zero was counted from opening the pump shutter (beginning of sample irradiation by pump beam). As representative data, time-evolution of probe deflection angle is plotted in Figs.~\ref{qd_timeevolution} and \ref{ns_timeevolution}, respectively, for colloidal QDs and NSs in toluene, at a few selected pump powers. Considering a zoom-in factor of approximately two between the vertical axes in Figs.~\ref{qd_timeevolution} and \ref{ns_timeevolution}, one can readily infer that the strength of the nonlinear response is significantly larger in QDs compared to that of NSs. Within the same dimensionality of the samples (0D QDs or 2D NSs), CsPbBr$_3$ samples show slightly larger nonlinear response as compared to the corresponding CsPbBr$_{1.5}$I$_{1.5}$ samples. 

As for the general behaviour of the time-evolution data, the probe beam deflection increases slowly and monotonically with time at low pump powers, to  reach a saturation in about 500~ms. With increasing pump power, the deflection begins to increase at a faster rate, attains a maximum and then decreases to some extent, before settling to a lower steady value. At moderate powers the region of maximum is broad and rate of decrease from maximum is slow. At higher powers, the peak is narrower, and the amount and rate of decrease from maximum are enhanced. The final steady value of deflection increases with increasing pump power. It is interesting to note the changing rate of rise and decay of the deflection angle with increasing power, respectively, before and after attaining the maximum. This indicates observation of nonlinear behaviour of heat transport at short length and times scales, decided respectively, by the focused laser spot size and the thermal diffusion time constant.     
     
The initial rise of the probe deflection angle with pump irradiation time is attributed to the formation of thermal lens due to nonuniform heating of the sample by the Gaussian pump laser. In this part, the heat transport is dominated by the thermal diffusion. The decrease in probe deflection after reaching a maximum, at higher pump powers, may be attributed to the onset of convective heat flow. Convection sets in when the temperature gradient crosses a threshold value. Once convection sets in, it enhances the heat transport and decreases the temperature gradient, resulting to a reduced beam deflection. 

A detailed modelling of the observed phenomena may involve solving the heat transport equation along with the hydrodynamic equations, taking account of the thermal diffusion and convection.~\cite{Karimzadeh2013, Schaertl1999, Karimzadeh2012} It may also need to include appropriate nonlinear terms in the heat transport~\cite{Goodman1964} to account for the changing rate of thermal lens formation due to diffusion, and decay due to convection. This is beyond the scope of this paper. 

In order to calculate the nonlinear refractive index, we read the maximum deflection angle from the time evolution data at different powers, and plot it as a function of pump power [Figs.~\ref{qd_timeevolution}(c) and \ref{ns_timeevolution}(c)]. The deflection angle increases linearly in the low (0--25~mW) and moderate (25--60~mW) power regimes, but the slopes are different in the two regimes. The sudden change in slope seems to be correlated with the onset of convection. For the pump power of 25~mW and higher, the effect of convection is perceivable in the time evolution data (Figs.~\ref{qd_timeevolution} and~\ref{ns_timeevolution}). Variation of the probe deflection angle with pump power shows signature of saturation at high powers, presumably  due to the saturation of absorption of light by the QDs and NSs. We fit the data in the two power regimes by straight lines of slightly different slopes, and calculate the nonlinear refractive index, $n_2$ from the slopes using Eq.~\eqref{max deflection}. The values of $n_2$ thus calculated are tabulated in Table~\ref{n2table}. These values of $n_2$ estimated from our measurements using near-resonant CW lasers on all-inorganic perovskite nanostructures are higher (due to the resonant enhancement) by several orders of magnitude as compared to those reported in the literature for organic-inorganic hybrid or all-inorganic perovskite nanostructures using off-resonant pulsed lasers.~\cite{Yi2017, Zhang2016, Mirershadi2016, Wang2015, Wu2018, Krishnakanth2018} For example, Lu {\etal} reports $n_2 \sim 10^{-17}$~m$^2$/W for CsPbBr$_3$ QDs in n-hexane from z-scan measurements using a pulsed laser (pulse width 130~fs and repetition rate 76~MHz) at $\lambda =800$~nm.~\cite{Lu2016}     

\begin{table}[htb]
    \caption{Values of nonlinear refractive index, $n_2$, at low and moderate powers.}
    \begin{tabular}{|c|c|c|c|}
        \hline
        \multicolumn{2}{|c|}{\multirow{2}{*}{Sample}} & \multicolumn{2}{c|}{$n_2 \times 10^9$~m$^2$/W} \\ \cline{3-4} 
        \multicolumn{2}{|c|}{} & Low power & Moderate power \\ \hline
        \multirow{2}{*}{QDs} & CsPbBr$_3$ (S1) & $2.81 \pm 0.05$ & $1.29 \pm 0.04$ \\ \cline{2-4} 
        & CsPbBr$_{1.5}$I$_{1.5}$ (S3) & $2.21 \pm 0.05$ & $1.12 \pm 0.05$ \\ \hline 
        \multirow{2}{*}{NSs} & CsPbBr$_3$ (S2) & $1.21 \pm 0.02$ & $0.95 \pm 0.03$ \\ \cline{2-4} 
        & CsPbBr$_{1.5}$I$_{1.5}$ (S4) & $1.08 \pm 0.03$ & $0.59 \pm 0.04$ \\ \hline
    \end{tabular}
    \label{n2table}
\end{table}

\begin{figure}[htb]
	\includegraphics[clip,width=0.8\linewidth]{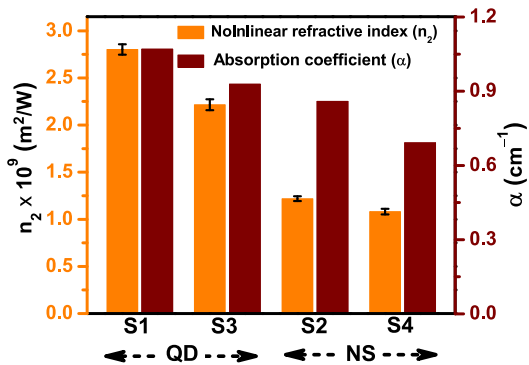}
	\caption{Bar plots comparing the nonlinear refractive indices ($n_2$) and absorption coefficients ($\alpha$) for different samples.} \label{comparison bar}
\end{figure}

Table~\ref{n2table} brings out the following trends: (a) nonlinear refractive index is significantly higher for the QDs compared to the NSs; and (b) mixed halide samples have lower nonlinear refractive index in comparison to the corresponding bromide samples. Consistent with this trend, the CsPbBr$_3$ QDs have the highest nonlinear refractive index while the CsPbBr$_{1.5}$I$_{1.5}$ NSs have the least. This is correlated with the absorption coefficient of these samples at $\lambda = 403$~nm. Samples having higher absorption coefficient is also possessing greater nonlinear refractive index [Fig.~\ref{comparison bar}]. This is consistent with the conclusion that the observed nonlinear optical effect is dominated by thermal lensing effect for liquid samples under irradiation by CW lasers. The vertical scales in Fig.~\ref{comparison bar} is chosen in such a way that the height of the bar for the nonlinear refractive index, $n_2$, and that for the absorption coefficient, $\alpha$, are nearly same for the CsPbBr$_3$ QDs, for which, both $n_2$ and $\alpha$ are the highest among all samples. This allows a comparison of the relative strengths of $n_2$ values in relation to the values of absorption coefficients, for other samples. While the height of the bars for refractive index are only about $50 - 60\,\%$ of that of the absorption coefficient for NSs, it is almost $90 - 100\,\%$ for the QDs. This suggests that besides the absorption coefficient, other thermal properties and viscosity of the sample solution are influenced by the dimensionality of the nanostructures.       

\begin{figure}[htb]
	\includegraphics[clip,width=0.8\linewidth]{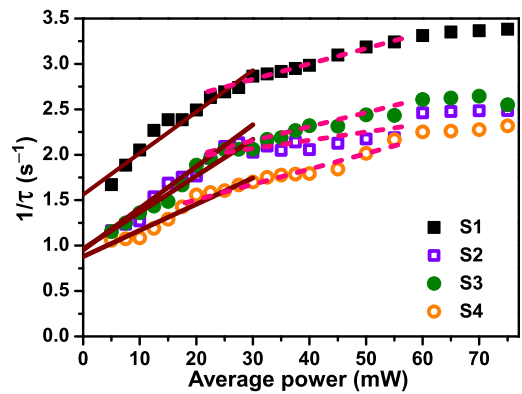}
	\caption{Variation of the heat diffusion rate with pump power for the QDs and NSs} \label{risetimes}
\end{figure}

In order to study the formation time of thermal lens, we define the time taken to reach the maximum deflection as the rise time, $\tau$, and define its inverse, $\tau^{-1}$, as the rate of heat diffusion. The heat diffusion rate thus calculated for all four samples are plotted as a function of pump power in Fig.~\ref{risetimes}. The heat diffusion rate increases almost linearly with the pump power in the low and moderate power regimes, though the rate of increase shows a sudden change at about 25~mW of pump power. This is correlated with the onset of convection, whose effect becomes prominent when pump power reaches 25~mW or higher. A saturation behaviour is seen at high powers. Within the linearised theory of heat diffusion,~\cite{Bergman2016} the thermal diffusion coefficient is given by $\alpha = K/(\rho  C_p)$, where $K =$~thermal conductivity, $C_p =$~specific heat capacity, and $\rho =$~density. The local density $\rho$ is determined by the local temperature arising due to the heating effect of the pump laser. But the local temperature at any given time during the pump heating is dependent on the heat transport rate. Thermal conductivity and specific heat capacity of the material may also depend on the local temperature. So the heat diffusion rate may not be a simple constant in this case. The local density at a given point within the pump beam spot decreases with increasing pump power due to increased temperature. This may lead to the observed increase of heat transport rate. 

At the end, we studied the dependence of nonlinear optical responses on concentration of the QDs and NSs in the colloidal solutions. While the time-resolved beam deflection data show very similar trends at different concentrations, all the observations scale up with increasing concentration. The convection effect shows up at a lower pump power. The slopes of maximum deflection angle and heat diffusion rate with pump power increase. Effective value of the nonlinear refractive index is larger for higher concentrations. But the sedimentation effect becomes prominent at higher concentrations. It was observed that  1~mg/ml is the optimum concentration at which we have carried out most of our experiments.

\section{Summary and conclusions}
In summary, we studied time-resolved nonlinear optical response of all-inorganic perovskites, CsPbX$_3$ (X~$=$~Br and Br:I in 50:50 stoichiometry) QD and NS colloids in toluene by a novel two-colour pump-probe beam-deflection technique using CW lasers. Exploiting the resonance nature, pump-beam excitation slightly above the absorption edges of the samples gives rise to significantly enhanced values of thermo-optical nonlinear refractive indices, which are larger by several orders of magnitude as compared to those reported in the literature for nonresonant measurements on similar or other perovskite nanostructures, using high power (amplified) femtosecond pulsed lasers. Heat diffusion rate was found to be dependent on laser intensity. This nonlinear heat transport may be related to the intricate relationship among thermal diffusion rate, local temperature, density, viscosity, and other thermo-optical coefficients. Onset of convective heat transport beyond a threshold intensity resulted to a nonmonotonic time-evolution of the beam deflection. These measurements may inspire further studies of heat transport in the micron length-scale and millisecond time-scale by a noncontact technique. The results may also draw interests of the theoreticians to develop models and computer simulations of micron-scale heat transport, taking account of convective flows and nonlinear relations among various thermo-optical coefficients. Due to their large thermo-optical nonlinearities, these materials, in conjunction with the beam-deflection technique using near-resonant CW diode lasers may find applications in fast optical modulators, display devices and scanning beam radar instruments.        

\begin{acknowledgements}
     SR thanks the Department of Science and Technology (DST), India, for INSPIRE fellowship; AM thanks the University Grants Commission (UGC), India, for his fellowship; SB thanks the financial support from DST-SERB under sanction no. EMR/2016/001703.
\end{acknowledgements}

\bibliography{beam_deflection_references}


\end{document}